\documentclass[%
 reprint,
nofootinbib,
 amsmath,amssymb,
 aps,
 prc,
floatfix,
twocolumn, superscriptaddress,
]{revtex4-2}
\usepackage[normalem]{ulem}
\usepackage{slashed}
\usepackage{graphicx}
\usepackage{dcolumn}
\usepackage{bm}
\usepackage{hyperref}

\hypersetup{
  colorlinks=true,    
  citecolor=blue,     
  linkcolor=blue,     
  urlcolor=blue       
}

\usepackage{orcidlink}
\usepackage{footnote}
\usepackage{hyperref}
\usepackage[x11names,svgnames,dvipsnames]{xcolor}
\usepackage{comment}

\newcommand{\pT}{p_T}
\newcommand{\xT}{x_T}
\newcommand{\mT}{m_T}
\newcommand{\meanpT}{\langle p_T\rangle}
\newcommand{\UxT}{U(x_T)}

\newcommand{\mupi}{\mu_\pi}
\newcommand{\snn}{\sqrt{s_{\rm NN}}}
\newcommand{\dd}{\mathrm{d}}

\newcommand{\WIR}{W_{\rm IR}}
\newcommand{\RBE}{R_{\rm BE}}
\newcommand{\AIR}{A_{\rm IR}}
\newcommand{\Tsw}{T_{\rm sw}}
\newcommand{\lnL}{\ln\mathcal{L}}
\newcommand{\hth}{\hat\theta_{\rm H}}
\newcommand{\GeVc}{GeV/$c$}

\usepackage{xcolor}
\usepackage{ulem}
\definecolor{purple}{rgb}{0.8,0,0.6}

\begin{document}

\title{Are hybrid hydrodynamic models missing soft-pion physics?}


\author{Osvaldo Ferreira\orcidlink{0000-0002-6711-8308}}
\email{osvaldof@if.usp.br}
\affiliation{
 Instituto de F\'isica, Universidade de S\~ao Paulo (USP), R. do Mat\~ao, 1371, S\~ao Paulo, Brazil, 05508-090
}%

\author{Thiago S. Domingues \orcidlink{0000-0001-6756-8600}}
\affiliation{
 Instituto de F\'isica, Universidade de S\~ao Paulo (USP), R. do Mat\~ao, 1371, S\~ao Paulo, Brazil, 05508-090
}%

\author{Fernando G. Gardim \orcidlink{0000-0002-4838-1469}}
\affiliation{Instituto de Ci\^encia e Tecnologia, Universidade Federal de Alfenas, 37715-400 Po\c cos de Caldas, MG, Brazil
}

\author{Cicero D. Muncinelli \orcidlink{0000-0003-0076-3852}}
\affiliation{Universidade Estadual de Campinas (Unicamp), R. S\'ergio Buarque de Holanda, 777, Campinas, Brazil, 13083-859
}

\author{Andr\'e V. Giannini
\orcidlink{0000-0003-0123-6472}}
\affiliation{
  Faculdade de Ci\^encias Exatas e Tecnologia, Universidade Federal da Grande Dourados, 
  Dourados, MS, Brazil, 79804-970
}
\affiliation{Departamento de F\'isica, Universidade do Estado de Santa Catarina, 89219-710 Joinville, SC, Brazil}

\author{Mauricio Hippert\,\orcidlink{0000-0001-5802-3908}}
\affiliation{Centro Brasileiro de Pesquisas F\'isicas, Rua Dr. Xavier Sigaud 150, Rio de Janeiro, RJ, 22290-180, Brazil}
\author{Rajeev Singh\,\orcidlink{0000-0001-5855-4039}}
\affiliation{Institut für Theoretische Physik, Goethe University, Max-von-Laue-Straße 1, D-60438 Frankfurt am Main, Germany}

\author{Gabriel S. Denicol \orcidlink{0000-0002-8395-9756}}
\affiliation{%
  Instituto de F\'isica, Universidade Federal Fluminense,
  Av. Milton Tavares de Souza, Niter\'oi, Brazil, 24210-346,
}%

\author{Tiago Nunes da Silva \orcidlink{0000-0001-8044-4055}}
\affiliation{%
 Departamento de F\'{i}sica, Centro de Ciências Físicas e Matemáticas, Universidade Federal de Santa Catarina, Campus Universit\'{a}rio Reitor Jo\~{a}o David Ferreira Lima, Florian\'{o}polis, Brazil, 88040-900}%

\author{Gabriel S. Rocha \orcidlink{0000-0001-7190-8137}}
\affiliation{%
  Instituto de F\'isica, Universidade Federal Fluminense,
  Av. Milton Tavares de Souza, Niter\'oi, Brazil, 24210-346,
}%

\author{Giorgio Torrieri \orcidlink{0000-0002-0611-766X}}
 \affiliation{%
  Universidade Estadual de Campinas (Unicamp), R. S\'ergio Buarque de Holanda, 777, Campinas, Brazil, 13083-859
}

\author{Jun Takahashi \orcidlink{0000-0002-4091-1779}}
 \affiliation{%
  Universidade Estadual de Campinas (Unicamp), R. S\'ergio Buarque de Holanda, 777, Campinas, Brazil, 13083-859
}

\author{Matthew Luzum \orcidlink{0000-0002-0367-7055}}
\affiliation{
 Instituto de F\'isica, Universidade de S\~ao Paulo (USP), R. do Mat\~ao, 1371, S\~ao Paulo, Brazil, 05508-090
}%

\collaboration{The ExTrEMe Collaboration}


\date{\today}
\begin{abstract}
In this work, we perform a systematic Bayesian analysis of the full parameter space of a state-of-the-art hybrid hydrodynamical model and show that no combination of parameters provides a satisfactory description of the low-$p_T$ region of the pion $(\pi^{\pm})$ transverse-momentum spectra. This suggests that an additional mechanism enhancing pion production at low $p_T$ may be needed to properly describe the ALICE data. We first reexamine and confirm the existence of an excess of low $p_T$ pions relative to model predictions, showing that no comparable excess is observed in the spectra of kaons $(K^\pm)$, protons $(p)$, or antiprotons $(\bar{p})$. We further demonstrate that this excess persists under a rigorous Bayesian analysis of simulations that include, e.g., a hadronic afterburner, a feature omitted in previous similar studies. Despite the observed excess, the scaled pion spectra retain their universal behavior, exhibiting little dependence on collision centrality. In fact, pions exhibit the most pronounced universality among the particle species considered. We then estimate the effects of two candidate mechanisms for generating the low-$p_T$ pion enhancement by introducing a one-parameter extension of the model into the Bayesian analysis. We find that the data favor enhancement mechanisms whose contributions decrease with increasing $p_T$, becoming negligible at high-$p_T$. Our results motivate further investigations of non-equilibrium effects in hybrid hydrodynamic simulations.
\end{abstract}

\maketitle


\section{Introduction}

Ultra-relativistic heavy-ion collisions create the conditions needed
to produce a quark-gluon plasma (QGP).
Over the past decades, relativistic viscous hydrodynamics has become
a central component of the theoretical description of its evolution~\cite{Heinz:2013th,Bernhard2019,JETSCAPE:2020mzn,Nijs2021,Martinoia:2024hip} and has been incorporated into the simulation chains used to model these collisions.
The precise measurements now available allow this description to be tested in detail, exposing discrepancies that may reveal missing effects from current simulations~\cite{Arslandok:2023utm}.

One such discrepancy is a persistent excess of low-$p_T$ pions in ALICE data~\cite{ALICE2013} compared with hybrid hydrodynamical simulations. Reference~\cite{Lu2025} performed a systematic study of this excess. Their hydrodynamical simulations combined TRENTo~\cite{Moreland2015}, for initial conditions, FluiduM~\cite{Floerchinger:2018pje} for hydrodynamic evolution, and FastReso~\cite{FastReso2019} for introducing the effects of resonance decays. Using this model, their Bayesian inference analysis scanned a six-dimensional parameter space (an overall normalization constant, the peaks in the shear and bulk viscosity to entropy ratio parameterizations, the initial fluid time, and the two freeze-out temperatures). Moreover, they assumed a sudden isothermal freeze-out and did not include a hadronic afterburner. This raises the question of whether the missing rescattering and the associated resonance regeneration could explain the excess, or if there is an additional effect that can enhance pion production.

Several enhancement mechanisms have been proposed to explain the low-$\pT$ excess, also called the \emph{soft-pion puzzle}~\cite{Lu2025,Coc:2026xfe}. As an example, Refs.~\cite{Kataja1990,GavinRuuskanen1991,Bebie1992,Hung1998,HiranoTsuda2002, Begun2014,Begun2015,Begun:2016cva, Blaschke:2020afk,Vitiuk2026} suggest that since pion-number-changing reactions stop before elastic ones, the pions can acquire an effective chemical potential $\mupi$ after chemical freeze-out. Other effects may come from the finite widths of the resonances that feed-down into pions~\cite{Huovinen2017, Coc:2026xfe}, a perturbative corona contribution~\cite{Kanakubo:2019ogh}, in-medium modifications of the pion dispersion relation~\cite{Fuchs:1996pa}, partially coherent emission~\cite{Ru:2017nkc}, or pion production from a gluon-saturated initial state~\cite{Zhao2019}. Other types of mechanisms connect the excess to chiral symmetry restoration and non-equilibrium effects associated with its critical dynamics. Near the chiral crossover, the pions pole mass and velocity drop below the vacuum values~\cite{Pisarski:1996mt}, which enhances the soft-pion yield already in equilibrium~\cite{Son:2001ff,Son:2002ci}. If the rapid expansion in HIC is fast enough, it can, in addition, drive them out of equilibrium~\cite{Florio:2025lvu,Florio2025, Bruschke2025}, an idea that goes back to the disoriented chiral condensates formed after a quench~\cite{Rajagopal:1993ah,Bjorken:1997re,Scavenius:1999zc}.

In this work, we perform a Bayesian analysis
using the simulated data of Ref.~\cite{JETSCAPE:2020mzn},
and show that the discrepancy is specific to pions and remains even when resonance feed-down, hadronic rescattering, and dynamical kinetic decoupling are accounted for. We scan the full parameter space of their model, which has seventeen parameters. Compared with Ref.~\cite{Lu2025}, the extra parameters give more flexibility to the parametrizations of the shear and bulk viscosities and include the TRENTo~\cite{Moreland2015} parameters for the initial conditions, which we treat as free. Even after this full scan, the excess remains, which suggests that additional physics is needed to explain the low-$p_T$ range of the pion spectra. Moreover, we argue that taking the pion enhancement into account can resolve the anomaly reported in Ref.~\cite{Domingues:2026mis}, where different calibrations of the Bayesian analysis lead to incompatible posteriors.

We investigate what type of modification in the spectra of produced particles is needed to reconcile hybrid hydrodynamical models with data. To fully address this problem, some of the mechanisms for generating the pion enhancement discussed above need to be implemented in the simulation chain. Moreover, to perform a Bayesian analysis, a large data set of simulated events is needed. 

Before performing such an expensive numerical analysis, we first perform a simpler one. The idea is that, under some simplifying assumptions, we can mimic the effect of a pion enhancement mechanism with a one-parameter template, i.e., a fixed $p_T$ shape
whose amplitude is a free parameter, acting on our baseline model. The strongest assumption is that this template can be applied to the final spectra, even tough it mimics a process that should take place at particlization. We therefore do not aim to fully implement the underlying mechanisms responsible for the pion enhancement. Instead, we seek to estimate their effect and establish an upper bound on its magnitude. As we will see, we can learn about the \emph{type} of modification required. Rigorous quantitative implementation of pion enhancement mechanisms is left for future work.


\begin{figure*}[ht]
    \centering
    \includegraphics[width=\linewidth]{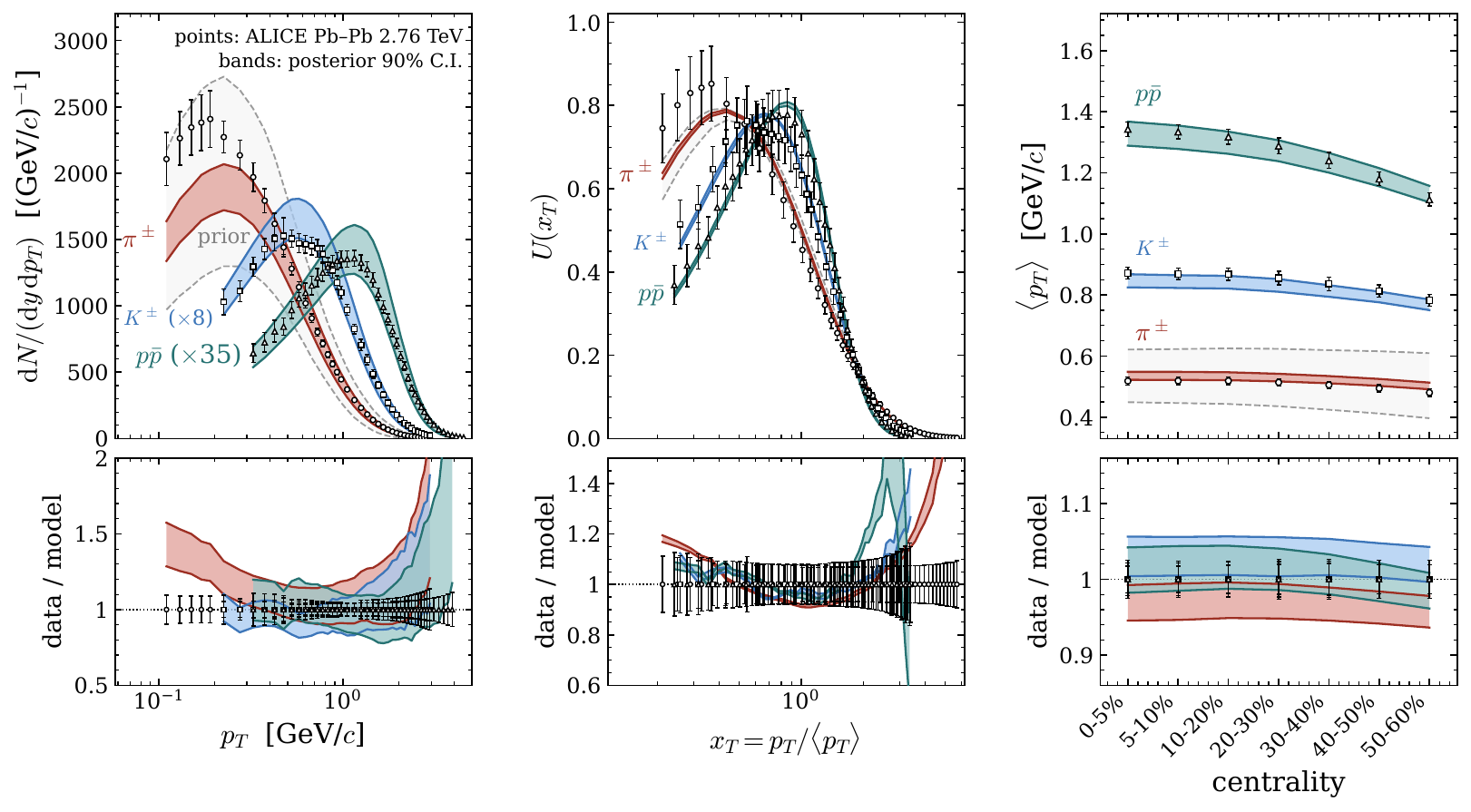}
    \caption{Left: absolute spectra $\dd N/(\dd y\,\dd\pT)$ in 0--5\% centrality. Middle: scaled spectra $\UxT$ in 0--5\% centrality, predicted. Right: $\meanpT$ in the seven centrality classes, predicted. Points are ALICE data~\cite{ALICE2013}, with corresponding errors. Bands are posterior 90\% credible intervals. Lower panels: data/model, with the data uncertainties drawn about unity. The hybrid hydrodynamical model underpredicts the low-$p_T$ region of the pion transverse momentum spectra. The figure shows that this effect is observed on the data/model ratios of both the absolute and scaled spectra. The same behavior is not observed for kaons and protons.
    }
    \label{fig: pion excess and other particles panel}
\end{figure*}


\section{The baseline model: hybrid hydrodynamic simulations}
\label{sec: baseline}

The baseline physics model employed in this work is the hybrid chain of the JETSCAPE/SIMS analysis \cite{JETSCAPE:2020mzn} and the Bayesian analysis methods will follow Ref.~\cite{Domingues:2026mis}. The chain is composed of: TRENTo initial conditions~\cite{Moreland2015}; a free-streaming pre-equilibrium stage~\cite{Broniowski2009,Liu2015}; (2+1)D viscous hydrodynamics with temperature-dependent shear and bulk viscosities, solved with MUSIC~\cite{Schenke:2010nt,JETSCAPE:2020mzn}; particlization at a switching temperature $\Tsw$ with the Cooper--Frye prescription~\cite{CooperFrye}, including the Grad 14-moment viscous correction $\delta f$~\cite{JETSCAPE:2020mzn,McNelis2021}; and the SMASH hadronic afterburner~\cite{SMASH2016}, which performs resonance decays, hadronic rescattering and dynamical kinetic decoupling. Of the particlization schemes studied in Ref.~\cite{Domingues:2026mis}, only the Grad $\delta f$ is used here. The chain has seventeen parameters which are listed in Table~\ref{tab:params}. The simulations were run at 500 design points. We use the same 485 points as Ref.~\cite{JETSCAPE:2020mzn}.

We compare the ALICE spectra of $\pi^\pm$, $K^\pm$, and $p+\bar p$ in Pb--Pb collisions at $\snn=2.76$~TeV~\cite{ALICE2013}, in seven centrality classes from 0--5\% to 50--60\%. Only points with transverse kinetic energy $m_T-m<2.364$~GeV are used, where $m_T=\sqrt{\pT^2+m^2}$ is the transverse mass and $m$ is the hadron mass. For pions, this means $\pT\le2.45$~\GeVc. This was done because at high $p_T$ the hydrodynamical description breaks down, and forcing the model to describe this region could spoil the calibration. The lowest measured $\pT$ is 0.11, 0.225, and 0.325~\GeVc\ for $\pi$, $K$ and $p$, respectively. This leaves 733 data points: 252 for pions, 237 for kaons, and 244 for protons. In the Bayesian inference framework, we closely follow Ref.~\cite{Domingues:2026mis}, using Gaussian-process emulators~\cite{KennedyOHagan2001,Higdon2008,Bernhard2016,JETSCAPE:2020mzn} trained on the design points after a Principal Component Analysis. Only the absolute spectra $dN/(dy\,dp_T)$, i.e., the yields per unit rapidity and transverse momentum, enter the likelihood directly. The other observables are \emph{predictions} of the trained emulators.

\section{The pion excess and the spectra of identified charged particles}
\label{sec:pion-excess}

Our discussion will focus on three observables that characterize the transverse momentum distribution of the identified charged particles: the absolute particle spectra $dN/(dy\,d\pT)$\footnote{We do not adopt the common ``differential'' spectra terminology to avoid confusion.}, the scaled particle spectra $\UxT \equiv \frac{\meanpT}{N}\,\frac{dN}{d\pT}$ with $\xT \equiv \pT/\meanpT$~\cite{ExTrEMe:2024fxt, Domingues:2026mis, Domingues:2026xvs}, and the mean transverse momentum $\meanpT$, which is a proxy for $\pT$-integrated observables. Here $N=\int d\pT\,(dN/d\pT)$ is the total particle yield. The scaled spectra carry information about the \emph{shape} of the distribution, $\meanpT$ about its momentum \emph{scale}, and the absolute spectra about both. 

For both the model and the data, we compute $\meanpT$ by integrating over the measured ALICE range. The scaled spectra, in contrast, are defined through integrals over the whole $\xT$ range. For the data, this requires an extrapolation of the measured range, which we do with a Blast-Wave fit.

Reference~\cite{Domingues:2026mis} examined the pion spectra and showed that calibrating the model with only integrated observables, such as $\meanpT$, leads to systematic discrepancies with the scaled spectra. Conversely, calibrating the model with $U(x_T)$ leads to an overestimation of $\meanpT$ and its fluctuations. From now on, we refer to this problem as the \emph{calibration tension}. Part of this tension is expected, since the two types of observables carry complementary information. We therefore calibrate the model with the absolute spectra, which contain both. Moreover, their analysis raises the question of whether this calibration tension is present only in the pion spectra.

Figure~\ref{fig: pion excess and other particles panel} shows the results for the three observables and their corresponding data-to-model ratios. At low $\pT$, the model underpredicts both the absolute and scaled $\pi^{\pm}$ spectra. In addition, as noted in Ref.~\cite{Domingues:2026mis} and shown in the middle column, the prior predictions for the pion spectra do not cover the data. Such a discrepancy is not observed for $K^{\pm}$, $p$, and $\Bar{p}$. These results agree with Ref.~\cite{Lu2025}, which did not include effects such as hadronic rescattering, the regeneration of resonances, resonance decays with finite lifetimes during the hadronic evolution, and a dynamical kinetic decoupling~\cite{SMASH2016}. Including these effects is therefore not enough to explain the soft region of the pion spectra.

\begin{figure}
    \centering
    \includegraphics[width=\linewidth]{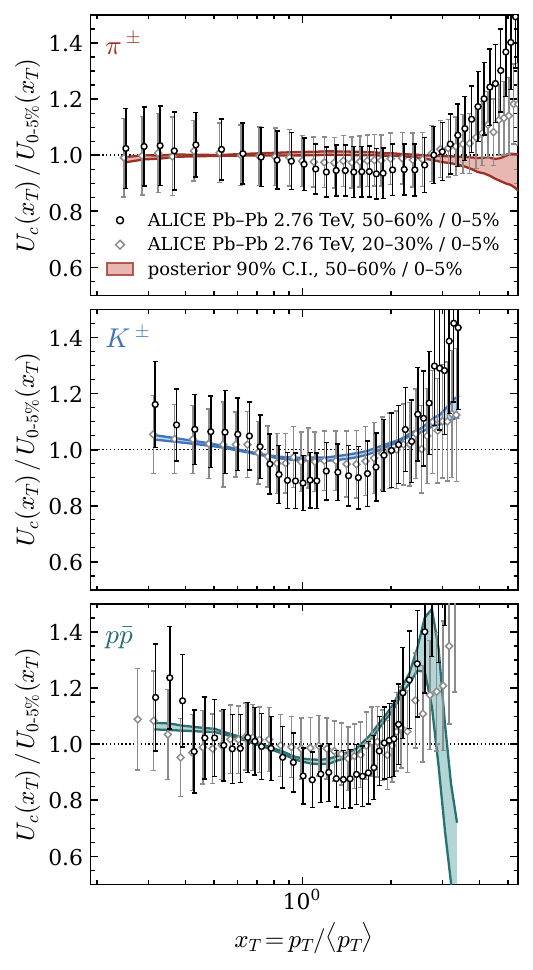}
    \caption{Centrality ratios $U_c/U_{0-5\%}$ as a function of $x_T$. The increase in $U_c/U_{0-5\%}$ ratios show that heavier particles produce a scaled spectra with stronger dependence with centrality.}
    \label{fig: Universality of particle species}
\end{figure}

In Fig.~\ref{fig: Universality of particle species}, we examine whether the underprediction of pions by the model can spoil the universality property of the scaled spectra. We find that it does not, and that the pion spectra is in fact the most universal across centrality classes. The deviation from universality is quantified by the centrality ratios $U_c/U_{0-5\%}$, where $U_c$ refers to the spectra of a given centrality class $c$. This deviation seems to increase with mass and is largest for protons. Reference~\cite{Guo:2026kqv} attributes this mass dependence to the Cooper-Frye prescription. 

\section{Pion enhancement mechanisms and their impact on the Bayesian analysis}
\label{sec: mechanisms}

In this section, we argue that if we take the pion enhancement into account and calibrate the model with the absolute spectra, we can better describe the pion spectra and resolve the calibration tension. We introduce two \emph{templates} for how the final pion spectra would be modified at low $\pT$. With them, we address two questions: \emph{could an enhanced pion spectrum solve the calibration tension? If so, can we estimate the enhancement and its behavior?}
\begin{figure*}[ht]
    \centering
    \includegraphics[width=\linewidth]{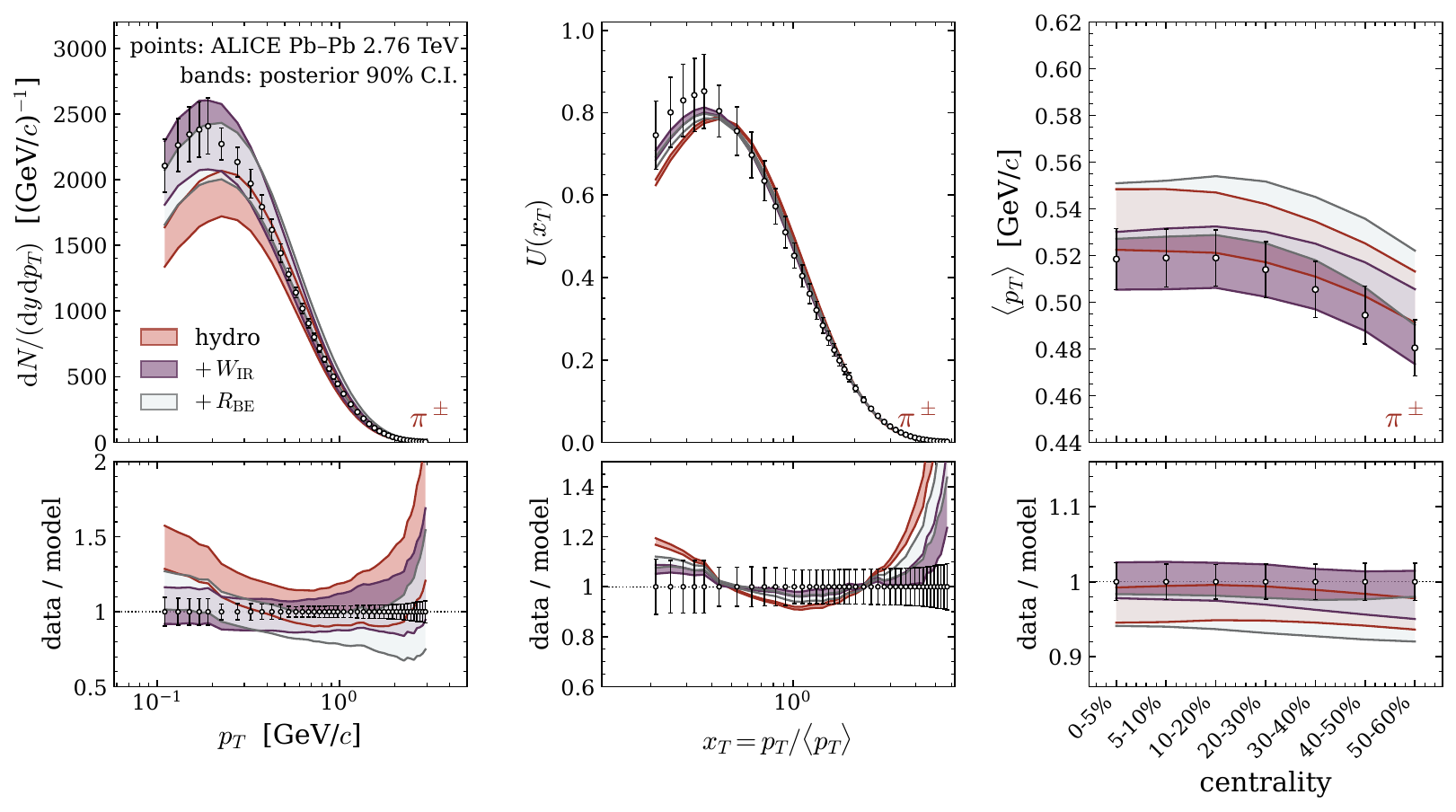}
    \caption{Pion absolute and scaled spectra and $\meanpT$ for the three calibrations: hydro only (red), with the infrared weight $\WIR$ (purple) and with the Bose--Einstein weight $\RBE$ (grey). Bands and points mean the same as in Fig.~\ref{fig: pion excess and other particles panel}. }
    \label{fig:mechanisms}
\end{figure*}

The core idea of this part of the analysis is to assume that we can mimic the effect of the underlying pion enhancement mechanisms by multiplying the final spectra by a template $W(\pT;\phi)$ as 
\begin{equation}
  \frac{\dd N'_\pi}{\dd y\,\dd\pT}=W(\pT;\phi)\,\frac{\dd N_\pi}{\dd y\,\dd\pT},
  \label{eq:Wspectrum}
\end{equation}
and then recalibrating the model to obtain the posteriors for the $17+1$ parameters (see Sec.~\ref{sec: recalibration}).
Each template $W\ge1$ depends on one extra parameter $\phi$ and is built such that $W = 1$ at $\phi=0$. The $\pT$-dependence of each template is fixed by the expected effect of their corresponding mechanism on the final spectra, which will be described in Secs.~\ref{sec: mu pi} and~\ref{sec: chiral}.

Before moving on, let us discuss some simplifying assumptions involved in the use of Eq.~\eqref{eq:Wspectrum}.
Both mechanisms below change how pions with a given momentum are produced when the fluid is converted to particles. In the Cooper-Frye prescription, the final pion spectra is obtained by summing the pions emitted by all fluid cells at the switching temperature $\Tsw$. This corresponds to an integration of $f=f_{\rm eq} + \delta f$ where $f_{\rm eq} = [\exp((u\cdot p)/\Tsw) - 1]^{-1}$, $u^{\nu}$ is the fluid velocity, and $\delta f$ is the viscous correction. 
When using Eq.~\eqref{eq:Wspectrum} we assume that the mechanisms act as $f\rightarrow W f$ and that $W$ can be factored out of the integration over fluid cells at $\Tsw$, i.e., it acts on all of them in the same way. Moreover, the final spectra are also affected by resonance decays and the hadronic afterburner, which take place after particlization. Therefore, assuming that they retain their $\pT$-dependence, our analysis can only quantify the remaining enhanced pions after these processes have taken place.

The analysis we perform with the templates should therefore be read as only estimates of the size and, above all, of the $\pT$ \emph{shape} of the modification that the data require. Despite this, we will show that this simplified analysis can reveal interesting information regarding the soft pion puzzle.

\subsection{The effective pion chemical potential}
\label{sec: mu pi}

The fast expansion of the QGP in heavy-ion collisions may take the system out of equilibrium. Different proposals suggest that this can generate an effective pion chemical potential through the loss of chemical equilibrium, either with a partial~\cite{GavinRuuskanen1991} or a full \cite{Kataja1990, Bebie1992, Begun2014, Blaschke:2020afk,Vitiuk2026} thermalization of pions. In both cases, pion number-changing processes, such as $\pi\pi\leftrightarrow\pi\pi\pi\pi$\footnote{Notice that resonance decays, such as $\rho\to\pi\pi$,
do change the pion number, however, the combination
$n_\pi+2n_\rho+3n_\omega+\dots$ is conserved~\cite{Bebie1992}.} are too slow compared to the expansion of the system, which prevents its chemical equilibration before freeze-out. The excess of pions produced by this mechanism generates an effective pion chemical potential $\mupi$.

In this work, we model the presence of this chemical potential through a template that re-weights the spectrum with a ratio of Bose--Einstein distributions with and without $\mupi$, both at the switching temperature and for a fluid cell at rest:
\begin{equation}
  R_{\rm BE}(\pT;\mupi)=
  \frac{e^{\mT/\Tsw}-1}{e^{(\mT-\mupi)/\Tsw}-1}.
  \label{eq:wBE}
\end{equation}

At large $\pT$, $\RBE\to e^{\mupi/\Tsw}$, so the chemical potential leaves a \emph{multiplicative offset} that survives to the high-$\pT$ tail. At $\pT\to0$, $\RBE$ saturates at $[e^{m_\pi/\Tsw}-1]/[e^{(m_\pi-\mupi)/\Tsw}-1]$. As $\mupi\to m_\pi$ this value diverges, which is the threshold for Bose--Einstein condensation~\cite{Begun2015}. The prior on $\mupi$ therefore stops at 0.135~GeV. SMASH already builds up pion chemical potentials dynamically during the hadronic stage. The fitted $\mupi$ therefore measures the \emph{residual} over-occupation relative to the calibrated chain, and, because feed-down pions dilute it, it is a lower bound on that of the primordial pions.

\subsection{Enhancement from critical chiral dynamics}
\label{sec: chiral}

The second template refers to the enhancement that may come from the critical dynamics of chiral symmetry breaking. If the system passes through the chiral crossover fast enough, the long-wavelength pion modes cannot follow and fall out of equilibrium: they are \emph{quenched}. References~\cite{Florio:2025lvu,Florio2025} combined a scaling analysis with stochastic simulations of the Model-G dynamics~\cite{HohenbergHalperin1977} of the chiral order parameter. They found that during the coarsening that follows the quench, the pion correlator is enhanced with respect to equilibrium by a factor 
\begin{equation}
  \frac{G_{\pi\pi}(t,k)}{G^{\rm eq}_{\pi\pi}(k)}
  \;\sim\;
  \begin{cases}
    1/(k\xi), & k\gg m_\pi,\\[2pt]
    1/(m_\pi\xi), & k\ll m_\pi,
  \end{cases}
  \label{eq:florio12}
\end{equation}
where $k$ is the pion momentum and $\xi$ is the chiral correlation length, and time-dependent scaling function of order one has been dropped. We use this expression to motivate the following template
\begin{equation}
  W_{\rm IR}(\pT;\AIR)=1+\AIR\,\frac{m_\pi}{\mT},
  \label{eq:wIR}
\end{equation}
which tends to $1+\AIR$ for $\pT\ll m_\pi$ and to $1+\AIR\,m_\pi/\pT$ for $\pT\gg m_\pi$. The amplitude $\AIR$ absorbs the factor $1/(m_\pi\xi)$ and the fraction of the enhancement that survives the hadronic stage. Unlike the chemical potential, which acts as a fugacity on the whole spectrum, this enhancement \emph{switches off} at large $\pT$. This distinction turns out to be crucial for our analysis.

\subsection{Bayesian recalibration}
\label{sec: recalibration}

Since both templates act as a multiplicative map on the baseline prediction, the extended model, with one additional parameter $\phi=\mupi$ or $\phi=\AIR$, is the composition
\begin{equation}
  y(\theta,\phi) \;=\; \mathcal{M}_\phi\big[\,y_0(\theta)\,\big].
  \label{eq:composition}
\end{equation}
Here $y_0(\theta)$ is the emulator prediction of the hybrid model at the 17 model parameters $\theta$, and $\mathcal{M}_\phi$ multiplies its pion spectra by $W(\pT;\phi)$, Eq.~\eqref{eq:Wspectrum}. The joint posterior is thus
\begin{equation}
  p(\theta,\phi\,|\,D)\;\propto\;
  p\big(D\,\big|\,\mathcal{M}_\phi[y_0(\theta)]\big)\;
  p(\theta)\,p(\phi),
\end{equation}
where $D$ denotes the data. It is sampled in the $(17+1)$-dimensional space, with uniform priors $\AIR\in[0,6]$ and $\mupi\in[0,0.135]$~GeV (Table~\ref{tab:params}).
Both priors start at 0, when we recover the baseline model. The upper egde of $\mupi$ is the pion mass. Above this value, pions would form a pion condensate \cite{Mannarelli:2019hgn}, and the modeling of this phase is out of the scope of this work. For $\AIR\in[0,6]$ the upper edge represents an enhancement of about $7$ times the expected by the baseline model, far above the observed excess and therefore preventing artifacts from the choice of the upper edge value.

From now on, we denote the three calibrations as hydro, $+\WIR$, and $+\RBE$. They share the data, the emulator, and the prior on $\theta$. Their predictions for the pion observables are compared in Fig.~\ref{fig:mechanisms}.

\section{Discussion}
\label{sec:discussion}

Both enhancement templates produce a better fit to the differential and scaled spectra, as shown in Fig.~\ref{fig:mechanisms}. However, the rightmost panel of Fig.~\ref{fig:mechanisms} shows that the critical dynamics template $\WIR$ produced a better result for $\meanpT$, resolving the calibration tension. This suggests that a simple one-parameter extension of the model can accommodate the excess of low-$p_T$ pions. However, the transverse-momentum dependence of the enhancement also matters.

In our analysis all that can be stated is comparative and qualitative. Nonetheless, we quote some of the main values that were obtained in the recalibration of the model with the one-parameter extension. More details of the model parameters can be found in Table~\ref{tab:extra} and the full posteriors are given in Table~\ref{tab:params} and Figs.~\ref{fig:corner-is} and \ref{fig:corner-hy}. 

We first discuss the infrared amplitude, for which we find a best-fit value of $\AIR=0.640$.
In Ref.~\cite{Domingues:2026mis}, it was shown that the nucleon width $w$ is very sensitive to the calibration tension. With the enhancement introduced by $\AIR$, the posterior median of the nucleon width moves from $w=1.42$~fm in the hydro-only calibration to $w=1.09$~fm, close to the value
$w\approx1.1$~fm obtained by JETSCAPE from $\pT$-integrated observables~\cite{JETSCAPE:2020mzn}. This can offer some intuition on why this mechanism resolves calibration tension. At the best fit, the number of model parameters lying at an edge of their prior
range drops from $11$ to $2$ to when compared to the hydro-only.
The resulting pion $\meanpT$ agrees with the ALICE values, with the model differing from the data by
at most $1.5\%$.

For the pion chemical potential, the best fit is $\mupi=37.3$~MeV. Again, the posterior median of the nucleon width moves to $w=0.97$~fm, slightly below the JETSCAPE value. We also observe the same reduction of model parameters lying at the edge of the priors. The most relevant difference is the prediction for the pion $\meanpT$, which is in this case overpredicted in all seven centrality classes, by $3.8-5.3\%$, i.e., by $1.5-2.1$ times the ALICE uncertainty.

In Appendix~\ref{App: parameters correlations}, we argue that the new parameters $\mupi$ and $\WIR$ are directly responsible for most of the fit improvement. In other words, although the recalibration with $\AIR$ and $\WIR$ does allow the Hydro model parameters to relax, this is only responsible for a minor fraction of the fit improvement. Performing the exercise of keeping the parameters fixed at their baseline most likely values, we show that $\WIR$ alone delivers 88\% of the total likelihood improvement of $W_{\rm IR}$, while for $R_{\rm BE}$ this number is 68\%.
In Appendix~\ref{App: parameters correlations} we also discuss how the new parameters correlate to the Hydro model parameters. Moreover, although not explicitly shown here, we have checked that the universality property of the scaled spectra is not altered by the pion enhancement templates.
 
The fitted amplitude can be roughly related to the chiral correlation length. For $k\ll m_\pi$, Eq.~\eqref{eq:florio12} gives an enhancement of order $1/(m_\pi\xi)$, so that $\AIR\simeq C/(m_\pi\xi)$, where $C$ is a number of order one\footnote{$C$ that depends on time; in the simulations of Ref.~\cite{Florio2025} it is between about 1 and 2.3 while the enhancement is
largest.}. However, only the pions produced directly at particlization, the \emph{primordial} pions, should be enhanced.
About half of the soft pions come instead from resonance decays~\cite{Coc:2026xfe}, and these are not enhanced.
To produce the same excess, the primordial pions must therefore be enhanced about twice as much, $A_{\rm prim}\simeq2\AIR\approx1.3$.
With $C=1$ this gives $m_\pi\xi\approx0.8$, i.e., $\xi\approx1.1$~fm, comparable to the pion Compton wavelength $1/m_\pi\simeq1.4$~fm, as expected when chiral symmetry is explicitly broken~\cite{Florio:2025lvu}.

The chiral dynamics template can also be discussed in terms of the effect proposed by Refs.~\cite{Son:2001ff,Son:2002ci}, which showed that near the chiral crossover, the pion pole mass $m_{0}$ and velocity drop below their vacuum values, already in equilibrium. Consider pions emitted with a pole mass $m_{0}=m_\pi-\Delta m$ and the vacuum velocity. Expanding $\sqrt{p^2+m_0^2}\simeq m_T-m_\pi\Delta m/m_T$, with $\Delta m\ll m_\pi$, and then the exponential, which requires $(\Delta m/T)(m_\pi/m_T)\ll1$, we find that the ratio of their Boltzmann factor to the vacuum one is 
\begin{equation}
  \frac{e^{-\sqrt{p^2+m_{0}^2}/T}}{e^{-\sqrt{p^2+m_\pi^2}/T}}\simeq1+\frac{\Delta m}{T}\,\frac{m_\pi}{m_T},
  \label{eq:SS}
\end{equation}
which is exactly the form of Eq.~\eqref{eq:wIR}, with $\AIR=\Delta m/T$. At leading order, therefore, our fit cannot tell the two scenarios apart. Taken literally, the best fit requires a large mass drop. Matching the soft limit, $e^{\Delta m/T}=1+\AIR$, gives $\Delta m\simeq T\ln(1+\AIR)\approx80$~MeV at $T=\Tsw\approx157$~MeV, beyond the linear regime of Eq.~\eqref{eq:SS}. For comparison, lattice QCD found that the pion quasiparticle mass drops from 130~MeV in vacuum to 81(2)~MeV at $T=154$~MeV~\cite{Krasniqi:2024kwm}, i.e., by about 40\%. With the same matching this would give $\AIR\approx0.4$, of the same order as the fitted value. 

Although this estimates should not be taken as measurements of the correlation length, they serve as indication that there is a rich phenomenology of non-equilibrium phenomena and chiral symmetry breaking to be explored in future work.

Finally, we mention some effects not treated here but that can also contribute to the soft pion enhancement. Within a different framework, Refs.~\cite{Coc:2026xfe, Huovinen2017} have considered the effects of thermal broadening on the production of light particles, and although such effects enhanced the production of pions, it was still not enough to explain the observed data. 
Species-dependent viscous corrections at particlization, derived from a
generalized relaxation-time approximation, also modify the identified-hadron
spectra and the $K/\pi$ and $p/\pi$ ratios, and the effect survives the hadronic
cascade~\cite{Aguiar:2026ioh}.
Our baseline uses the Grad $\delta f$ only, so such corrections are not
included here. We also notice that Ref.~\cite{SanMartin:2023zhv} has studied the effect of an extended hadron list, within a blast-wave estimate with fixed parameters, and have shown that it leaves the multiplicity-normalized pion spectrum practically unchanged after hadronic rescattering, shifting the pion $\meanpT$ by about 1\%.

\section{Conclusions}

We have shown that the state-of-the-art hybrid hydrodynamical model~\cite{JETSCAPE:2020mzn, Domingues:2026mis} fails to explain the low-$p_T$ region of the pion transverse-momentum spectra. The model underpredicts the lowest-$\pT$ pions ($\pT=0.11$~\GeVc) by 43\% [29, 57]\% in 0--5\% collisions, with the 90\% credible interval in brackets, and we have shown that no similar discrepancy is observed for kaons or protons. Despite this discrepancy, it does not spoil the universality of the scaled spectra of pions, i.e., its  independence of the centrality class. In fact, pions show the most universal behavior, while the deviation from universality increases with mass~\cite{Guo:2026kqv}.

We then show that a one-parameter extension of the model, allowing for an enhancement of the production of soft pions, can significantly improve the fit of the pion spectra. Although qualitative, our analysis reveals that to simultaneously have a consistent description of $\meanpT$, \emph{the shape of the enhancement matters.} Both mechanisms add one parameter,  but $\WIR$ provides a better description of the observables compared to $\RBE$. The difference comes from $\pT\gtrsim0.5$~\GeVc. While $\RBE$ leaves a multiplicative factor $e^{\mupi/\Tsw}\approx1.25$ at high $\pT$, $\WIR$ fades as $m_\pi/m_T$. This indicates that the data favor a mechanism that increases soft pion production but does not distort the spectrum at larger transverse momentum.

Although both enhancement templates are responsible for most of the improvement in the fit, they also remove strain from the hydro parameters. A prominent example is the nucleon width, $w$, which gains some freedom to move from the upper end of the prior ($\sim 1.5$~fm) and relaxes to $\approx1.0$~fm, the value favored by calibrations to $\pT$-integrated observables~\cite{Domingues:2026mis, JETSCAPE:2020mzn}. This suggest that the lack of an enhancement mechanism for pion production is a possible source of the calibration tension~\cite{Domingues:2026mis}.

Our results suggests that current simulations of hybrid hydrodynamical models might be missing important physics regarding the production of soft pions. A number of theoretical proposals exist in this direction~\cite{Florio2025, Florio:2025lvu, Son:2001ff, Begun2014, Begun2015, Vitiuk2026} and its incorporation in the simulation chains deserves consideration. This would address some of the main limitations of our approach, and would make the discussion of the physical parameters related to chiral symmetry breaking and pion dynamics quantitative.

\begin{acknowledgments}
We acknowledge the use of the large language models
Claude Fable~5, Claude Opus~5.1, and Claude Opus~5.5 (Anthropic)
to extend the Bayesian analysis code of Ref.~\cite{Domingues:2026mis}. All code changes were reviewed and validated by the authors, who take full responsibility for the results. This study was financed, in part, by the São Paulo Research Foundation (FAPESP), Brasil within the FAPESP Thematic Grant 2023/13749-1 and the postdoctoral fellowship with Process Number 2026/03908-3. T.S.D. acknowledges financial support from the Brazilian National Council for Scientific and Technological Development (CNPq) through a PhD fellowship under process No.\ 141432/2025-0, INCT-FNA grant number 464898/2014-5, and from the Brazilian Federal Agency for Support and Evaluation of Graduate Education CAPES Foundation scholarship under process No. 88881.220538/2025-01. A.V.G. acknowledges support from CNPq through the Universal Grant 405458/2025-8.
MH is supported by the Brazilian National Council for Scientific and Technological Development (CNPq) under process No. 313638/2025-0.
A.V.G., M.H., G.S.R., G.S.D., F.G., O.F. and T.N.dS. acknowledge support from CNPq through the INCT-FNA grant 408419/2024-5. RS acknowledges the support of the Humboldt research fellowship for postdocs from the Alexander von Humboldt Foundation.  F.G. was supported by CNPq through 307806/2025-1 and 405458/2025-8.
\end{acknowledgments}

\begin{table*}[!t]
\caption{Model parameters and prior ranges. The parametrization is that of
Ref.~\cite{JETSCAPE:2020mzn}. The 17 model parameters are shared by the three
calibrations; $\AIR$ and $\mupi$ exist only in their mechanism's calibration. The last
three columns give the posterior median and 90\% credible interval (5th--95th
percentile) in each calibration. Temperatures are in GeV.}
\label{tab:params}
\footnotesize
\setlength{\tabcolsep}{3pt}
\begin{ruledtabular}
\begin{tabular}{lllccc}
symbol & meaning & prior & hydro & $+\WIR$ & $+\RBE$ \\ \hline
Norm & initial-density normalization & $[10,20]$ & 16.3 [13.9, 19.6] & 16.7 [12.8, 19.5] & 14.5 [12.0, 18.3] \\
$p$ & TRENTo generalized-mean parameter & $[-0.7,0.7]$ & $-0.35$ [$-0.67$, 0.33] & $-0.38$ [$-0.68$, 0.42] & $-0.19$ [$-0.63$, 0.35] \\
$\sigma_k$ & nucleon-fluctuation width & $[0.3,2.0]$ & 1.18 [0.55, 1.83] & 1.45 [0.53, 1.96] & 1.24 [0.52, 1.89] \\
$w$ & nucleon width [fm] & $[0.5,1.5]$ & 1.42 [0.90, 1.50] & 1.09 [0.59, 1.43] & 0.97 [0.66, 1.36] \\
$d^3_{\min}$ & min.\ inter-nucleon distance$^3$ [fm$^3$] & $[0,4.913]$ & 3.0 [0.8, 4.7] & 2.6 [0.8, 4.4] & 2.4 [0.5, 4.4] \\
$\tau_R$ & free-streaming time [fm/$c$] & $[0.3,2.0]$ & 0.74 [0.36, 1.17] & 0.65 [0.31, 1.33] & 0.64 [0.31, 1.42] \\
$\alpha$ & power of $\tau_{\rm fs}$ in energy density & $[-0.3,0.3]$ & $-0.14$ [$-0.29$, 0.02] & $-0.01$ [$-0.28$, 0.23] & 0.12 [$-0.22$, 0.28] \\
$T_\eta$ & temperature of the $\eta/s$ kink & $[0.13,0.30]$ & 0.261 [0.148, 0.298] & 0.241 [0.156, 0.296] & 0.203 [0.134, 0.293] \\
$a_{\rm low}$ & $\eta/s$ slope below the kink [GeV$^{-1}$] & $[-2,1]$ & $-1.2$ [$-2.0$, 0.9] & $-1.0$ [$-1.9$, 0.9] & $-0.4$ [$-1.9$, 0.9] \\
$a_{\rm high}$ & $\eta/s$ slope above the kink [GeV$^{-1}$] & $[-1,2]$ & $-0.06$ [$-0.93$, 1.38] & 0.21 [$-0.72$, 1.59] & 0.55 [$-0.72$, 1.76] \\
$(\eta/s)_{\rm kink}$ & $\eta/s$ at the kink & $[0.01,0.20]$ & 0.157 [0.065, 0.198] & 0.083 [0.017, 0.192] & 0.142 [0.034, 0.198] \\
$(\zeta/s)_{\max}$ & height of the $\zeta/s$ peak & $[0.01,0.20]$ & 0.110 [0.021, 0.172] & 0.088 [0.012, 0.173] & 0.067 [0.012, 0.183] \\
$T_\zeta$ & temperature of the $\zeta/s$ peak & $[0.12,0.30]$ & 0.226 [0.196, 0.271] & 0.270 [0.189, 0.299] & 0.249 [0.194, 0.298] \\
$w_\zeta$ & width of the $\zeta/s$ peak & $[0.025,0.15]$ & 0.039 [0.026, 0.104] & 0.049 [0.026, 0.129] & 0.053 [0.026, 0.122] \\
$\lambda_\zeta$ & asymmetry of the $\zeta/s$ peak & $[-0.8,0.8]$ & 0.62 [$-0.41$, 0.79] & 0.61 [$-0.63$, 0.79] & 0.62 [$-0.51$, 0.79] \\
$b_\pi$ & shear relaxation time [$\eta/(sT)$] & $[2,8]$ & 4.9 [2.5, 7.6] & 4.8 [2.6, 7.0] & 5.8 [2.8, 7.8] \\
$\Tsw$ & switching temperature & $[0.130,0.165]$ & 0.1527 [0.1466, 0.1608] & 0.1574 [0.1512, 0.1630] & 0.1593 [0.1539, 0.1643] \\ \hline
$\AIR$ & $\WIR$ strength, Eq.~\eqref{eq:wIR} & $[0,6]$ & --- & 0.691 [0.576, 0.800] & --- \\
$\mupi$ & $\RBE$ chem.\ potential, Eq.~\eqref{eq:wBE} [GeV] & $[0,0.135]$ & --- & --- & 0.0361 [0.0299, 0.0419] \\
\end{tabular}
\end{ruledtabular}
\end{table*}

\appendix

\section{Details of the model and the Bayesian analysis}
\label{App: Bayesian details}

Our analysis closely follows Ref.~\cite{Domingues:2026mis}, whose Secs.~II and~III describe the hybrid model, the emulators and the sampling. The general framework is that of Ref.~\cite{Bernhard2016} (Secs.~III\,B and III\,C) and the parameters are listed in Table~\ref{tab:params}. Here we summarize only what differs from those works and what is needed to understand the figures.

\paragraph*{Observables and emulator.} Instead of the pion scaled spectra, we fit the absolute spectra of pions, kaons and protons jointly (Sec.~\ref{sec: baseline}). The logarithm of the spectra is projected onto six principal components common to all species and centrality classes, each emulated by a Gaussian process. The weights of Sec.~\ref{sec: mechanisms} act on the emulator output, so the three calibrations share one emulator, and the extended models reduce exactly to the baseline at zero strength. Experimental uncertainties are treated as uncorrelated (as is the standard practice~\cite{Domingues:2026mis, Bernhard2016}), and only the Grad $\delta f$ is considered.

\paragraph*{Likelihood.} As in Refs.~\cite{Bernhard2016,JETSCAPE:2020mzn}, the likelihood is a multivariate normal,
\begin{equation}
  -2\lnL = \chi^2 + \ln\det\Sigma, \qquad \chi^2 = r^{\sf T}\Sigma^{-1}r,
  \label{eq:chi2}
\end{equation}
with $r=y-m(\theta,\phi)$ the difference between data and emulator prediction, and $\Sigma$ the sum of the experimental and emulator covariances, including the truncated principal components. Since $\Sigma$ depends on the parameters, we use $-2\lnL$ to compare different fits.

\begin{figure*}[!t]
    \centering
    \includegraphics[width=\linewidth]{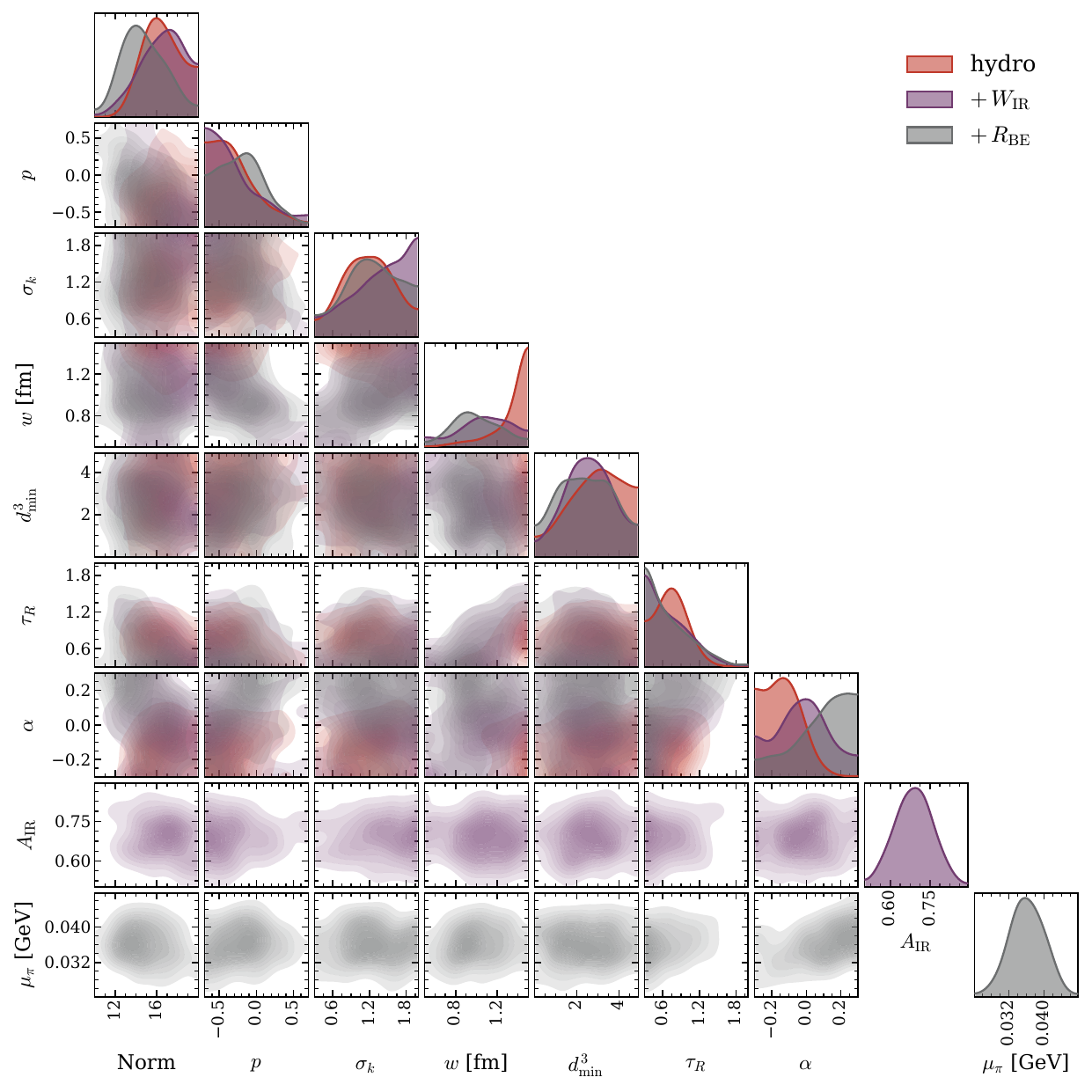}
    \caption{Marginal one- and two-dimensional posteriors of the initial-state and free-streaming parameters (Table~\ref{tab:params}), plus $\AIR$ and $\mupi$, for the hydro-only (red), $+\WIR$ (purple) and $+\RBE$ (grey) calibrations. The ($\mupi$, $\AIR$) cell is empty by construction, since no calibration contains both.}
    \label{fig:corner-is}
\end{figure*}

\begin{figure*}[h]
    \centering
    \includegraphics[width=\linewidth]{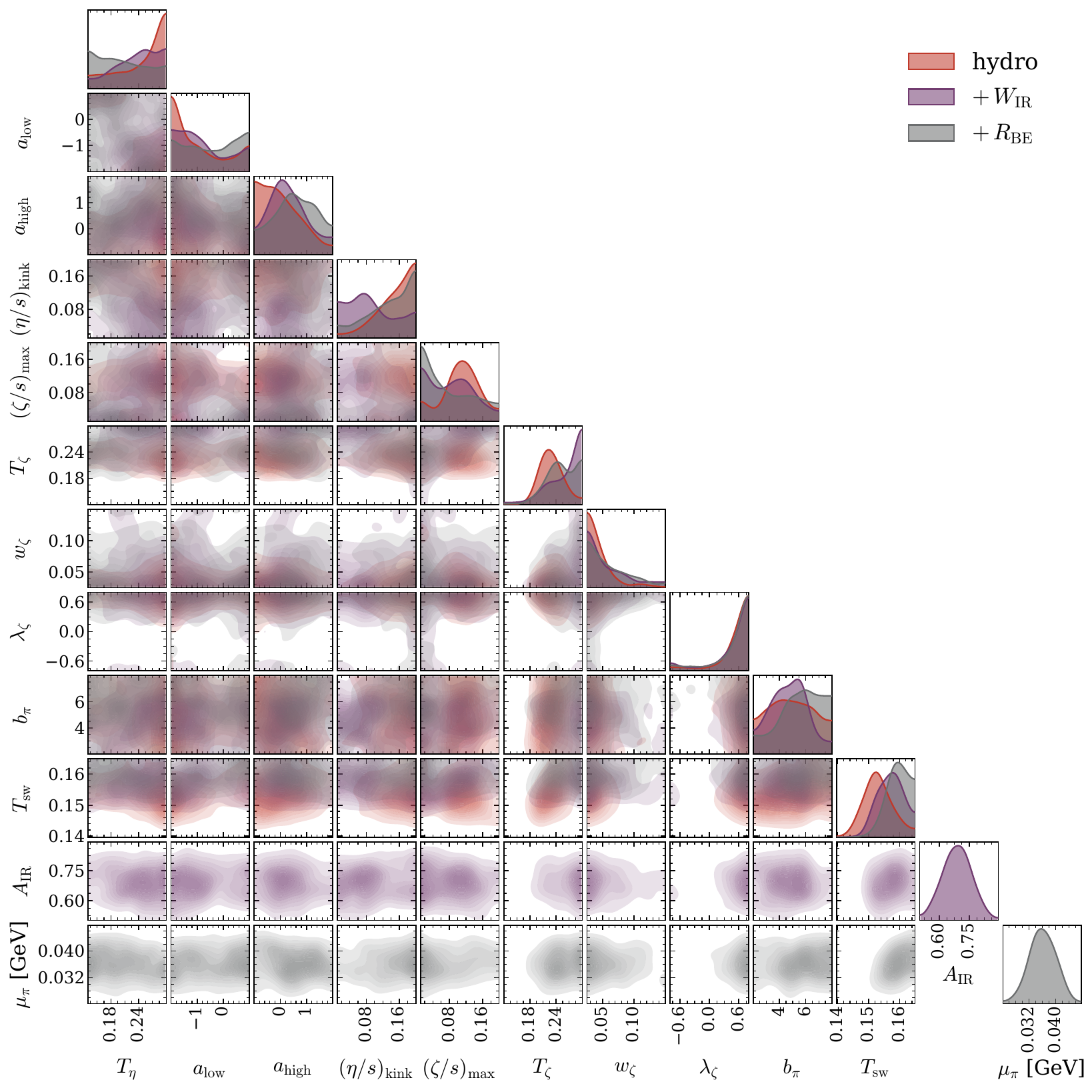}
    \caption{Marginal one- and two-dimensional posteriors of the hydrodynamic and particlization parameters, plus $\AIR$ and $\mupi$.}
    \label{fig:corner-hy}
\end{figure*}

\paragraph*{Physics parameters.} Figures~\ref{fig:corner-is} and~\ref{fig:corner-hy} compare the three posteriors. Spectra alone imposes only weak constraints on viscosities, which might be expected since flow observables were not used in the calibration~\cite{Bernhard2019,JETSCAPE:2020mzn}. Without a soft-pion enhancement, several parameters are pushed against the edges of the prior, while both enhancement templates relax some of these parameters. The largest changes occur in $w$, $\Tsw$ and the free-streaming and bulk-viscosity parameters, which are the parameters that control the shape of $\UxT$~\cite{Domingues:2026mis}.

\begin{table}[h]
\caption{Estimates of the extra parameters: best fit with profile-likelihood intervals, and posterior median with 90\% credible interval.}
\label{tab:extra}
\begin{ruledtabular}
\begin{tabular}{lcc}
 & $\AIR$ & $\mupi$ [MeV] \\ \hline
best fit & 0.640 & 37.3 \\
profile 68\% & [0.614, 0.672] & [35.0, 39.7] \\
profile 90\% & [0.577, 0.703] & [33.2, 41.2] \\
posterior median & 0.691 & 36.1 \\
posterior 90\% & [0.576, 0.800] & [29.9, 41.9] \\
\end{tabular}
\end{ruledtabular}
\end{table}

\section{How do the one-parameter extensions correlate with the model parameters?}\label{App: parameters correlations}

In this section, we discuss how the enhancement template parameters correlate to the model parameters and how much of the fit improvement is delivered by the new parameters $\AIR$ and $\mupi$, and not only by the relaxation of the other parameters.

\begin{figure*}
  \centering
  \includegraphics[width=0.86\textwidth]{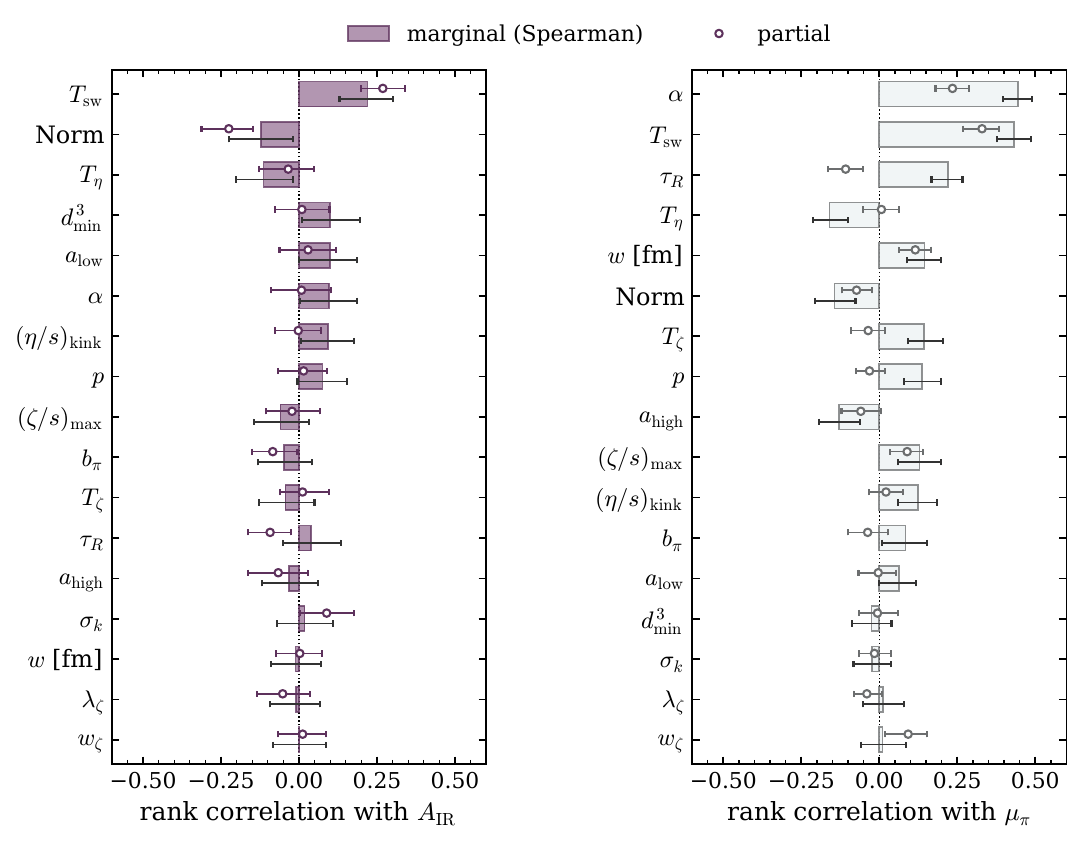}
  \caption{Posterior correlation of the extra parameter with each of the 17 physics
  parameters of Table~\ref{tab:params}: $\AIR$ in the $W_{\rm IR}$ calibration (left)
  and $\mupi$ in the $R_{\rm BE}$ calibration (right). Filled bars show the Spearman
  rank correlation $\rho_s$, with the black error bar giving its 90\% bootstrap
  interval. Open circles show the partial correlation $\rho_{\rm part}$, with its own
  90\% interval. Rows are sorted by $|\rho_s|$. }
  \label{fig:corr}
\end{figure*}

\paragraph*{Correlations.} We follow Ref.~\cite{Auvinen2020} and use the Spearman rank correlation coefficient $\rho_s$~\cite{Spearman1904} to quantify the correlation between the new parameters and the hybrid hydrodynamical model parameters. This coefficient is designed to capture any monotonic relation between two parameters, and the results are shown in Fig.~\ref{fig:corr}. The figure also shows the partial correlation $\rho_{\rm part}$, which removes the contribution given by the other 16 parameters~\cite{Whittaker1990}. Error bars represent the uncertainties regarding the  size of these correlations. We find that $\AIR$ is nearly independent of the physics parameters. Its strongest correlation is with $\Tsw$ and the normalization, which, like the weight, increases the pion yield. Together, all 17 parameters predict only 15\% of its variance. For $\RBE$, $\mupi$ is more strongly tied to the hydrodynamic parameters (33\%), having its strongest correlation with $\alpha$ and $\Tsw$. This is expected since $\Tsw$ is used in the definition of $\RBE$ in Eq.~\eqref{eq:wBE}.

\begin{figure*}[h]
  \centering
  \includegraphics[width=0.86\textwidth]{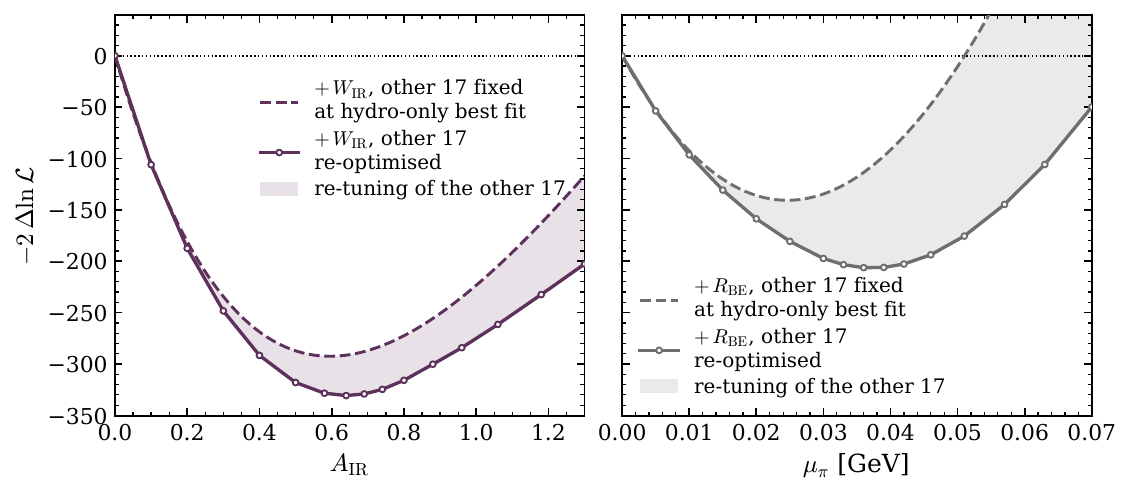}
  \caption{$-2\Delta\lnL$, Eq.~\eqref{eq:dlnL}, as a function of
  the extra parameter: $\AIR$ (left) and $\mupi$ (right). Dashed: the conditional curve
  $-2\Delta\lnL(\hth,\phi)$, with the 17 physics parameters frozen at $\hth$. Solid with
  points: the profile curve $\min_\theta[-2\Delta\lnL(\theta,\phi)]$, with the 17 re-optimized at each
  grid point. The shading between them is the gain from re-tuning. Lower is better. On
  the right, the dashed curve continues above the axis for $\mupi\gtrsim0.055$~GeV.
  }
  \label{fig:prof}
\end{figure*}

\paragraph*{Mechanism versus re-tuning.} We measure every fit relative to the hydro-only best fit $\hth$,
\begin{equation}
  -2\Delta\lnL(\theta,\phi) \equiv -2\,\bigl[\lnL(\theta,\phi)-\lnL(\hth,0)\bigr],
  \label{eq:dlnL}
\end{equation}
and Fig.~\ref{fig:prof} shows two cuts of it as functions of $\phi=\mupi, \AIR$. The \emph{conditional} curve, $-2\Delta\lnL(\hth,\phi)$, keeps the physics parameters frozen and measures the fit improvement that comes from the new parameter alone. The \emph{profile} curve, $\min_\theta[-2\Delta\lnL(\theta,\phi)]$, re-optimizes them at each $\phi$~\cite{Cowan2011}. The gap between the two is the gain from re-tuning.

For $\WIR$ the two curves nearly coincide. The mechanism alone delivers almost 90\% of the improvement, and its preferred amplitude barely moves. $\RBE$ behaves differently. On its own it saturates at a small $\mupi$, and above about 50~MeV it fits worse than the baseline. Reaching its best fit requires re-tuning, mainly of $\alpha$ and $\Tsw$, which supplies about a third of the gain. Since $\Tsw$ ends at the upper edge of its prior, this values might be underestimated.


\clearpage

\bibliographystyle{apsrev4-1}
\bibliography{references.bib}

\end{document}